\documentclass[aps,preprint,showkeys,showpacs,amssymb]{revtex4}
\usepackage{amsmath,amsthm}

\begin{document}

\newtheorem{thm}{Theorem}
\newtheorem{lem}[thm]{Lemma}
\newtheorem{claim}[thm]{Claim}

\newcommand\ket[1]{\ensuremath{|#1\rangle}}
\newcommand\bra[1]{\ensuremath{\langle#1|}}
\newcommand\iprod[2]{\ensuremath{\langle#1|#2\rangle}}
\newcommand\oprod[2]{\ensuremath{|#1\rangle\langle#2|}}

\title{Comparability of multipartite entanglement}
\author{Zhengfeng Ji}
  \email{jizhengfeng98@mails.tsinghua.edu.cn}
\author{Runyao Duan}
  \email{dry02@mails.tsinghua.edu.cn}
\author{Mingsheng Ying}
\affiliation{
State Key Laboratory of Intelligent Technology and Systems,\\
Department of Computer Science and Technology,\\
Tsinghua University,\\
Beijing 100084, P. R. China}

\begin{abstract}
We prove, in a multipartite setting, that it's always feasible to exactly transform a genuinely $m$-partite entangled state with sufficient many copies to any other $m$-partite state via local quantum operation and classical communication. This result affirms the comparability of multipartite pure entangled states.
\end{abstract}

\maketitle

\section{Introduction}\label{sec:intro}

Due to its numerous important applications\cite{BBC+93,BW92,DEJ+96,DEJ+98,BSST99} in the area of quantum information and computation, entanglement has been extensively studied in the past decade. The concept of considering entanglement as a new type of useful resource has been widely accepted. Efforts have been made to quantitatively and qualitatively measure this kind of resource in both bipartite and multipartite cases\cite{BBPS96,BPR+01,SM96}. In the bipartite case, a unique measure has been found\cite{BBPS96} to quantify entanglement asymptotically, however, it is a much more complicated problem for multipartite entanglements even in the asymptotic manner\cite{BPR+01,DVC00,Sudb00,BC01,Kemp99}.

Entanglement measure is in fact an approach of quantifying entanglement that relies closely on the reversibility of entanglement transformation. We assign a same scalar number to two different entangled states if they can convert interchangeably to each other by some predefined physical transformation. Because of the reversibility, we can simply do quantification by comparing the entanglement under consideration with some standard entanglements, for instance, the maximally entangled states or \textit{minimal reversible entanglement generating set} (MREGS)\cite{BPR+01}. However, the structure of entanglement is so complicated that it is difficult to define proper entanglement measure based on this interconvertible scheme except for the bipartite asymptotic case.

Various approaches can be adopted to cope with this difficulty. One common approach is to use multicomponent measures. For example, in the bipartite case, Nielsen's criteria of exact LOCC transformation\cite{Niel99} indicates that the vector of Schmidt coefficients, instead of any scalar, is a proper entanglement measure when exact transformations are considered. The disadvantage of this approach is that there exist incomparable states due to lack of total order on multicomponent measure\cite{BC01,Kemp99}. Another approach that we are concerned with is to measure entanglement relatively. In this approach, the requirement of reversibility is discarded and there are no so called standard entanglements. Instead, we focus on how many target entangled states can be obtained asymptotically, taking both initial state and target state as parameters. Specifically, define the \textbf{entanglement exchange rate} between $\ket{\psi}$ and $\ket{\phi}$ as\footnote{This definition is borrowed from the concept of exchange rate in finance and we adopt indirect quotation notation to make it compatible with the basic requirement of entanglement measure: to denote more entanglements with larger values}
\begin{equation}\label{eqn:rate}
R(\ket{\psi}, \ket{\phi}) = \sup \{ \frac{m}{n}\ :\ \ket{\psi}^{\otimes n}\text{ is convertible to }\ket{\phi}^{\otimes m}\}.
\end{equation}

One can indeed use entanglement exchange rate to define entanglement measure. For example, in the bipartite case and using asymptotic transformation, entanglement measure can be defined as
\begin{equation}
E(\ket{\psi}) = R(\ket{\psi}, \ket{\beta_{00}})
\end{equation}
while conversion in Eq(\ref{eqn:rate}) is specified in asymptotic manner. The entanglement exchange rate related to measure $E$ is then
\begin{equation}\label{eqn:re}
R_E(\ket{\psi}, \ket{\phi}) = \frac{E(\ket{\psi})}{E(\ket{\phi})}.
\end{equation}
One of the most different aspects between entanglement exchange rate and entanglement measure is that, for the latter one, the following very strong transitivity
\begin{equation}
R_E(\ket{\psi_0}, \ket{\psi_1}) \times R_E(\ket{\psi_1}, \ket{\psi_2}) = R_E(\ket{\psi_0}, \ket{\psi_2})
\end{equation}
always holds according to Eq(\ref{eqn:re}) while for general entanglement exchange rate we only have a weaker transitivity
\begin{equation}
R(\ket{\psi_0}, \ket{\psi_1}) \times R(\ket{\psi_1}, \ket{\psi_2}) \le R(\ket{\psi_0}, \ket{\psi_2})
\end{equation}
which is obvious since for most transformations of interest, their combination is still the same type of transformation. This inequality can be interpreted that, just as in economics, \textit{bid/offer spread} may also occur in entanglement exchange process.

To guarantee that the approach of entanglement exchange rate is not trivial, the quantity $R(\ket{\psi}, \ket{\phi})$ is required to be nonzero, and in turn this is equivalent to the assertion that it is always feasible to exactly transform an entangled states with sufficient many copies to any other entangled state. For the bipartite case, this assertion is just a simple corollary of the Nielsen's theorem\cite{Niel99} when exact LOCC transformation is considered. In order to warrant the approach of entanglement exchange rate also works in multipartite case, it is needed to verify the above assertion for $m$-partite states with $m \ge 3$. The difficulty for this purpose is that we have no longer the Nielsen's theorem and even the Schmidt decomposition for the multipartite case. In this paper, we are able to prove that for any $m \ge 2$, each $m$-partite entangled state with sufficient many copies can be transformed to any other $m$-partite state. And we organize this paper as follows. First, Entanglements that are \textbf{genuinely $m$-partite entangled} ($m$-partite entangled for short) are formally defined. Then we give the key lemma of this paper which correlates the multipartite and bipartite case. Finally, we prove the main result using the key lemma and conclude briefly.

\section{Comparability of pure entanglements}

Entanglements that are genuinely multipartite entangled are states that can not be written in product form between any bipartite partition of the parties. With $m=3$, for example, the $\ket{GHZ}$\cite{GHZ89} and $\ket{W}$ state are $3$-partite entangled while
\begin{equation}
\frac{\ket{001}+\ket{111}}{\sqrt{2}}
\end{equation}
is not if each of the three qubits belongs to a different party. We can formally describe this concept as follows.

Define $\mathcal{P}$ to be the set of all parties under consideration,
\begin{equation}
\mathcal{P} = \{P_i, 1\le i\le m\}.
\end{equation}
A state $\ket{\psi}$ shared by $m$ parties in $\mathcal{P}$ is called $m$-partite entangled state if for any nonempty proper subset of $\mathcal{P}$, say $\mathcal{A}$, $\ket{\psi}$ is entangled according to the bipartite partition $\mathcal{A}$ and $\mathcal{P}-\mathcal{A}$.

The elegant majorization criteria\cite{Niel99} indicates that, from sufficient many copies of bipartite entanglements, any other entanglements are always obtainable via LOCC only. Though there is no exact counterpart of this criteria in the multipartite case, Nielsen's theorem is still useful here. It allows us to concentrate bipartite entanglement to Bell states with which teleportation can be done between different parties. Teleportation back and forth in fact removes the constrains of LOCC and allows any global operations at cost of the consumption of shared entanglements.

Obviously, the main obstacle to generalize to the case of $m \ge 3$ is then how to convert the multipartite states to bipartite entanglement with which the above idea can be realized. The following lemma devotes to pad this gap.

\begin{lem}\label{lem:multi_to_bi}
Let $\ket{\psi}$ be a state shared by $n+m$ parties,
$$A_1, A_2, \cdots, A_n, B_1, \cdots, B_m.$$
Denote
\begin{eqnarray}
\mathcal{A} & = & \{A_i, 1\le i\le n \}, \nonumber\\
\mathcal{B} & = & \{B_j, 1\le j\le m \}. \nonumber
\end{eqnarray}
If \ket{\psi} is entangled between the partition $\mathcal{A}, \mathcal{B}$, there exist some $i, j$ and some LOCC procedure that produces a pure entangled state between $A_i$ and $B_j$.
\end{lem}

Two claims are given beforehand in the following to simplify the proof of Lemma \ref{lem:multi_to_bi}.

\begin{claim}\label{claim:phase_diff_only}
If there exists nonzero $\lambda$ such that $\ket{\phi_0}\ket{\phi_1}+\lambda\ket{\psi_0}\ket{\psi_1}$ is product state (without normalization), then $|\iprod{\phi_0}{\psi_0}|=1$ or $|\iprod{\phi_1}{\psi_1}|=1$, that is, $\ket{\phi_0}, \ket{\psi_0}$ or $\ket{\phi_1}, \ket{\psi_1}$ differ only in some global phase.
\end{claim}

\begin{proof}
Consider two projective measurements including
$$\oprod{\phi_0}{\phi_0}\otimes\mathcal{I}$$
and
$$\oprod{\psi_0}{\psi_0}\otimes\mathcal{I}$$
as one of the projective operators respectively. If one (the first, for example) of the projective operators occurs with zero probability in the corresponding measurement, then the norm of $\ket{\phi_0}(\ket{\phi_1}+\lambda\iprod{\phi_0}{\psi_0}\ket{\psi_1})$ is zero, which indicates that $\ket{\phi_1}, \ket{\psi_1}$ differ only in some global phase. Otherwise, since $\ket{\phi_0}\ket{\phi_1}+\lambda\ket{\psi_0}\ket{\psi_1}$ is product state, measurement on one party does not change the state of the other party. Thus there exists some nonzero complex number $c$ such that
\begin{equation}
\ket{\phi_1}+\lambda\iprod{\phi_0}{\psi_0}\ket{\psi_1} = c(\iprod{\psi_0}{\phi_0}\ket{\phi_1}+\lambda\ket{\psi_1}),
\end{equation}
that is,
\begin{equation}
(1-r\iprod{\psi_0}{\phi_0})\ket{\phi_1} = (c\lambda-\lambda\iprod{\phi_0}{\psi_0})\ket{\psi_1}.
\end{equation}
$\ket{\phi_1}, \ket{\psi_1}$ are not zero vector, so it's impossible that only one of $1-c\iprod{\psi_0}{\phi_0}$ and $c\lambda-\lambda\iprod{\phi_0}{\psi_0}$ is zero. If both of them are nonzero, $\ket{\phi_1}, \ket{\psi_1}$ differ only in some global phase. Or else both $1-c\iprod{\psi_0}{\phi_0}$ and $c\lambda-\lambda\iprod{\phi_0}{\psi_0}$ are zero, then we have
\begin{eqnarray}
\iprod{\psi_0}{\phi_0} & = & \frac{1}{c},\\
\iprod{\phi_0}{\psi_0} & = & c.
\end{eqnarray}
Thus,
\begin{equation}
|\iprod{\phi_0}{\psi_0}|^2 = \iprod{\psi_0}{\phi_0} \iprod{\phi_0}{\psi_0} = 1.
\end{equation}
In this case, $\ket{\phi_0}, \ket{\psi_0}$ differ only in a global phase.
\end{proof}

\begin{claim}\label{claim:one_prod}
$\ket{\phi}$ is bipartite entangled state and $\ket{\psi_0}\ket{\psi_1}$ is product state, then there is at most one nonzero real $\lambda$ such that $\ket{\phi}+\lambda\ket{\psi_0}\ket{\psi_1}$ is a product state.
\end{claim}

\begin{proof}
If there are two different nonzero real $\lambda$, say $\lambda_0, \lambda_1$, that satisfy the requirement, we have
\begin{eqnarray}
\ket{\phi}+\lambda_0\ket{\psi_0}\ket{\psi_1} & = & \ket{\alpha}\ket{\beta},\\
\ket{\phi}+\lambda_1\ket{\psi_0}\ket{\psi_1} & = & \ket{\alpha'}\ket{\beta'}.
\end{eqnarray}
Minus above two equations we get,
\begin{equation}
\ket{\alpha'}\ket{\beta'} = \ket{\alpha}\ket{\beta}+(\lambda_1-\lambda_0)\ket{\psi_0}\ket{\psi_1}.
\end{equation}
By Claim \ref{claim:phase_diff_only}, $\ket{\alpha}, \ket{\psi_0}$ or $\ket{\beta}, \ket{\psi_1}$ differ only in a global phase which contradicts with the fact that $\ket{\phi}$ is entangled.
\end{proof}

Assisted by above two claims, we are now ready to prove Lemma \ref{lem:multi_to_bi}.

\begin{proof}[Proof of Lemma \ref{lem:multi_to_bi}]
We consider projective measurement on $A_1$ and discuss two different cases separately. Let $\ket{\alpha_0}$ be a pure state of $A_1$ and
$$P=\oprod{\alpha_0}{\alpha_0}.$$
\textit{Case 1}. There exists some $P$ such that $(P\otimes\mathcal{I})\ket{\psi}$ is purely entangled between $\mathcal{A}'=\mathcal{A}-\{A_1\}$ and $\mathcal{B}$. Extend $\ket{\alpha_0}$ to a complete orthogonal basis in the $d$ dimensional state space of $A_1$, $\{\ket{\alpha_i}, 0\le i < d\}$, we have
$$\ket{\psi} = \sum_{i=0}^{d-1}\sqrt{p_i}\ket{\alpha_i}\ket{\psi_i},$$
where $\ket{\psi_0}$ is entangled between $\mathcal{A}', \mathcal{B}$. If there is some $\ket{\psi_i}$ not entangled between $\mathcal{A}', \mathcal{B}$, say $\ket{\psi_1}$, we can change the basis $\{\ket{\alpha_i}, 0\le i < d\}$ properly to $\{\ket{\alpha_i^{'}}, 0\le i < d\}$ and make the corresponding $\ket{\psi_0^{'}}, \ket{\psi_1^{'}}$ both entangled. Specifically, let
\begin{eqnarray}
\ket{\alpha_0^{'}} & = & \cos\theta\ket{\alpha_0} + \sin\theta\ket{\alpha_1},\\
\ket{\alpha_1^{'}} & = & \sin\theta\ket{\alpha_0} - \cos\theta\ket{\alpha_1},\\
\ket{\alpha_i^{'}} & = & \ket{\alpha_i}, 2\le i < d.
\end{eqnarray}
Then we have\cite{HJW93}
\begin{equation}\begin{pmatrix}
\sqrt{p_0^{'}}\ket{\psi_0^{'}}\\
\sqrt{p_1^{'}}\ket{\psi_1^{'}}\\
\end{pmatrix}=\begin{bmatrix}
\cos\theta & \sin\theta\\
\sin\theta & -\cos\theta\\
\end{bmatrix}\begin{pmatrix}
\sqrt{p_0}\ket{\psi_0}\\
\sqrt{p_1}\ket{\psi_1}\\
\end{pmatrix}.
\end{equation}
Employing Claim \ref{claim:one_prod}, it's easy to see that $\theta$ needed for our purpose exists. Continuously applying the technique of changing the basis, we can make all the $\ket{\psi_i}$ entangled. Thus $A_1$ can do a projective measurement according to the chosen basis, leaving an entangled pure state between $\mathcal{A}', \mathcal{B}$ after the measurement.\\
\textit{Case 2}. Otherwise, for any state $\ket{\alpha_0}$, $(P\otimes\mathcal{I})\ket{\psi}$ is product state. Then we can expand $\ket{\psi}$ as
\begin{equation}
\ket{\psi}=\sum_{i=0}^{d-1}\sqrt{p_i}\ket{i}\ket{\phi_i}\ket{\psi_i}.
\end{equation}
Setting $\ket{\alpha_0} = \ket{i}+\lambda\ket{j}$, we get
\begin{equation}
\sqrt{p_i}\ket{\phi_i}\ket{\psi_i}+\lambda\sqrt{p_j}\ket{\phi_j}\ket{\psi_j}
\end{equation}
is also a product state. Claim \ref{claim:phase_diff_only} states that for any $i, j$, $\ket{\phi_i}, \ket{\phi_j}$ or $\ket{\psi_i}, \ket{\psi_j}$ differ only in a global phase. Simple calculation tells that all states in $\{\ket{\phi_i}, 0\le i<d\}$ or in $\{\ket{\psi_i}, 0\le i<d\}$ differ only in a global phase. However the later case is impossible since it contradicts with the fact that $\ket{\psi}$ is entangled between $\mathcal{A}, \mathcal{B}$. In the former case, the reduced state of $\ket{\psi}$ on $A_1, B_1, B_2, \cdots, B_m$ is an entangled pure states.\\
In both cases above, we managed to reduce the number of parties without destroying the entanglement between $\mathcal{A}, \mathcal{B}$. Repeat the procedure several times, we will be able to find two parties $A_i$ and $B_j$ fulfilling our aim.
\end{proof}

Multiple uses of Lemma \ref{lem:multi_to_bi} lead to the main theorem of this paper:

\begin{thm}
For any $m$-partite entangled state $\ket{\psi}$ and any $m$-partite state $\ket{\phi}$, it is always feasible to obtain $\ket{\phi}$ provided that the available state $\ket{\psi}$ are sufficient and exact LOCC are allowed. Or we can abbreviate the result using entanglement exchange rate as $R(\ket{\psi}, \ket{\phi}) > 0$.
\end{thm}

\begin{proof}
Let $\mathcal{P}=\{P_i, 1\le i\le m\}$ be the $m$ parties and $\mathcal{A}=\{P_1\}$, $\mathcal{B}=\mathcal{P}-\mathcal{A}$. The proof of Lemma \ref{lem:multi_to_bi} indicates that there exists some $j\ne 1$ such that $\ket{\psi}$ is transformed to a bipartite pure entangled state between $P_1$ and $P_j$. The value of $j$ depends on the measurement result of each step. Different possible outcomes may lead to different $j$ and, of course, different type of entanglement between $P_1$ and $P_j$. But regardless of this indeterminism, sufficient many $m$-partite entanglements can produce sufficient many bipartite entanglements between at least one pair of parties, say $P_1$ and $P_{j^*}$. With these sufficient many bipartite entanglements (that belong to finite possible types) and teleportation, all operation on the joint system of $P_1$ and $P_{j^*}$ can be performed using only LOCC and thus, we can think of $P_1$ and $P_{j^*}$ as a single party. The same arguments work until the number of parties decreases to one. That is, among the $m$ parties, we can perform any quantum operations which include the preparation of another state $\ket{\phi}$.
\end{proof}

\section{Conclusions}\label{sec:conc}

In sum, we prove the comparability of genuinely $m$-partite entangled states in the sense of exact LOCC transformation. This affirms the nonzero property of entanglement exchange rate introduced in section \ref{sec:intro}. And it might be an indication that the approach of entanglement exchange rate is worth of studying since it simplifies the characterize of entanglement property. The mixed case extension of our result fails to hold because of the existence of bound entangled states\cite{HHH98}. This analysis tells that pure entangled states differ quantitatively while mixed entangled states differ qualitatively in our interpretation.

As an initiative attempt to study entanglement exchange rate, this work is preliminary and experimental. Many related problems need further investigations such as bounding entanglement exchange rate, calculating the exchange rate for some special states and analyzing the exchange rate by defining different type of convertibility. We hope that the research of entanglement exchange rate may help us to understand and interpret various phenomena in quantum information theory.

This work was partly supported by the National Foundation of Natural Sciences of China (Grant No: 60273003).

\bibliographystyle{apsrev}
\bibliography{come}

\end{document}